\documentclass[11pt]{article}

\usepackage{packages}
\usepackage{theorems}
\usepackage{notation}

\hypersetup{
    colorlinks=true,
    citecolor=blue,
    linkcolor=blue,
    urlcolor=blue
}

\hypersetup{pdftitle={Adaptivity Gaps for Influence Maximization}}

\title{Adaptivity Gaps for Influence Maximization: Percolation and Depletion}

\author{Gabriel de Azevedo\footnote{Cornell University. Email: gm587@cornell.edu} \  and Ulysse Hennebelle\footnote{Cornell Tech, Cornell University. Email: uh34@cornell.edu}}
\date{}

\begin{document}

\pagenumbering{roman}
\maketitle
\thispagestyle{empty}

\begin{abstract}
In the \emph{influence maximization} problem, a set of at most $k$ vertices, called \emph{seeds}, must be selected in a network where each arc is independently present with a known probability in order to maximize the expected number of vertices that are reached from the seeds.
A \emph{static policy} selects all seeds in advance, while an \emph{adaptive policy} selects them sequentially and draws on the feedback provided by previous cascades.
A major open problem is establishing the \emph{adaptivity gap} between these two classes.
First, for undirected graphs, we prove an adaptivity gap of at most $5\sqrt{k}$, where $k$ is the seed budget. This is the first nontrivial upper bound that holds independently of the graph. 
Our proof relies on two connection inequalities from percolation theory. 
Second, for general directed graphs on $n$ vertices, we obtain an $\mathcal{O}(\log n/\log\log n)$ bound, improving the previous $\lceil n^{1/3}\rceil$ guarantee. Third, we establish a constant bound of $2$ for bi-directed forests, unifying and  improving existing constant bounds for various orientations of forests.
These two results rely on a depletion approach that bounds the influence remaining after successive rounds of seed selection.
Finally, we improve the best known lower bound from $e/(e-1)$ to $1+\varphi$, where $\varphi$ is the golden ratio.
\end{abstract}

\clearpage
\pagenumbering{arabic}

\section{Introduction}

\emph{Influence maximization} is the problem of selecting a set of at most $k$ vertices, called \emph{seeds}, to maximize the expected spread of an activation cascade in a network.
The idea of influence maximization was developed by Domingos and Richardson~\cite{DomingosR01,RichardsonD02}, who distinguished the intrinsic and the network value of a customer in viral marketing. Since then, influence maximization has seen many applications both in practice and theory, such as viral marketing~\cite{hinz2011seeding,ChenWW10}, microfinance diffusion~\cite{banerjee2013diffusion}, and outbreak detection in networks~\cite{leskovec2007cost}.
In their seminal work, \citet*{KempeKT03} studied this problem under two standard diffusion models: the \emph{independent cascade model}, in which each arc is live independently with a probability given by a vector $\mathbf{p}$, and a seed activates every vertex reachable through live paths, and the \emph{linear threshold model}, where each vertex becomes active when the total weight of its active in-neighbors reaches a randomly drawn threshold.
For surveys of diffusion models, algorithms, and applications, see~\cite{LiFWT18,BanerjeeJP20}.
In this paper, we focus on the independent cascade model.

In the same work, \citet*{KempeKT03} showed that influence is a monotone submodular function of the seed set.
The classical analysis of greedy maximization therefore yields a $(1-1/e)$-approximation to the optimal set of $k$ seeds~\cite{NemhauserWF78}.
For undirected influence networks, in the independent cascade model, \citet*{KhannaL14} proved that the same greedy algorithm achieves a $(1-1/e+c)$-approximation for a universal constant $c>0$.
This improvement shows that undirected diffusion obeys a structure that permits stronger guarantees than monotone submodularity alone.

A \emph{static policy}, also called a \emph{non-adaptive policy}, chooses all seeds before observing the diffusion process.
In many applications, however, seeds can be \emph{probed} one at a time, so an \emph{adaptive policy} can use \emph{feedback} from earlier seeds to guide later choices. 
Adaptive influence maximization has, for example, been used to plan health campaigns, adapting participant selection to information revealed during earlier interventions~\cite{yadav2016using}.
The two main feedback models were introduced by \citeauthor{GolovinK11}~\cite{GolovinK10,GolovinK11}. Under \emph{full-adoption feedback}, the policy observes the entire cascade of each seed before choosing the next one.
They showed that this problem fits the framework of adaptive submodularity and thus that adaptive greedy achieves a $(1-1/e)$-approximation to the optimal adaptive policy.
Under \emph{myopic feedback}, the policy observes only the states of arcs leaving the selected seed.
Peng and Chen~\cite{PengC19} obtained a constant-factor approximation algorithm for influence maximization under this feedback model.
Our focus is on full-adoption feedback.

Approximation guarantees compare algorithms with the optimal policy for a given model, but do not determine how much feedback improves the optimum itself.  First formalized by \citet*{DeanGV04}, the \emph{adaptivity gap} measures, for a given model, how much the information given by the feedback can increase the optimal expected influence by measuring the ratio between the expected value of an optimal adaptive policy and that of an optimal static policy.
In the context of influence maximization, adaptivity gaps have already been studied heuristically in practice~\cite{jankowski2017balancing}, and in theory, first by \citet{ChenP19}, who established constant upper bounds for several restricted classes of directed graphs as well as a lower bound of $e/(e-1)$. Until now, this was the best known lower bound.
For general directed graphs on $n$ vertices, \citet*{DAngeloPV23} obtained the first sublinear upper bound, $\lceil n^{1/3}\rceil$.
For undirected graphs, they also proved a bound of $\sqrt{\alpha}+\mathcal{O}(1)$, where $\alpha$ is the sum of the degrees of vertices of degree greater than two.
Before the present work, no general upper bound depending only on the number of seeds $k$ improved on the elementary linear bound of $k$ implied by adaptive submodularity, even for undirected graphs.

This limited understanding contrasts with the well-established bounds for the related model of stochastic monotone submodular maximization.
For that setting, \citet{AsadpourN16} proved a tight adaptivity gap of $e/(e-1)$ under a matroid constraint.
This bound is permitted by the independence of item outcomes.
In influence maximization, cascades generated by different seeds share a realization of the graph and can therefore be dependent, so these independence-based bounds do not apply directly.

\subsection{Problem Definition}\label{sec:problem_definition}
An instance of \emph{influence maximization} consists of a finite set of vertices $V$ of cardinality $n$ and a probability vector $\mathbf{p}\in[0,1]^{V^2}$. 
Throughout, vectors are written in bold and their entries in plain type.
Each instance $(V,\mathbf{p})$ is associated with a random arc set $A\subseteq V^2$. For $A'\subseteq V^2$,
\[
    \Pr(A = A') = \prod_{a\in A'}p_a\prod_{a\in V^2\setminus A'}(1-p_a).
\]

We call an arc \emph{live} if it belongs to $A$.
Hence, every potential arc $a\in V^2$ is live independently with probability $p_a$. Unless otherwise specified, probabilities and expectations are taken over $A$ with the distribution given above and over any policy or sampling randomness present.

We define the \emph{cascade function} $C_V: 2^V \to 2^V$ by letting $C_V(S)$ be the set of vertices reachable from at least one vertex of $S \subseteq V$ along paths of live arcs. We call $C_V(S)$ the \emph{cascade} of $S$.
We say that a vertex in $C_V(S)$ is \emph{activated} by $S$.
The \emph{influence} of $S$ is $\sigma_V(S) = \card{C_V(S)}$, the number of vertices reached by $S$.
The function $\sigma_V$ is \emph{monotone}\footnote{A function $f:2^{V}\rightarrow\mathbb{R}$ is monotone if, for every $T\subseteq S\subseteq V$, $f(T)\le f(S)$.} and \emph{submodular}\footnote{A function $f:2^{V}\rightarrow\mathbb{R}$ is submodular if, for every $S,T\subseteq V$, $f(S\cap T)+f(S\cup T)\le f(S)+f(T)$.} and, by linearity, so is the function $S\mapsto\E[\sigma_V(S)]$~\cite{KempeKT03}.
When the vertex set $V$ is clear from the context, we omit it from the subscript.
When the set consists of a single seed $v$, we use $C(v)$ and $\sigma(v)$ as shorthand for $C(\set{v})$ and $\sigma(\set{v})$.

Given an instance $(V, \mathbf{p})$ and an integer $k \in \N$, we define the value of the best static policy as
\begin{equation*}
    \Static_k(V,\mathbf{p}) = \max_{S\subseteq V,\,|S| \leq k}\E[\sigma(S)].
\end{equation*}
We refer to this number $k$ as the \emph{budget} and to the set of vertices chosen by the policy as \emph{seeds}.

For an instance $(V,\mathbf{p})$, we define an \emph{adaptive policy} $\pi$ with budget $k$ to be a random sequence $(v_1, \ldots, v_k)$ of vertices in $V$ such that, for every $i \in[k]$, the choice of the vertex $v_i$ is a function of $C(v_1), \ldots, C(v_{i-1})$ and any external randomness independent of $A$.
We use $S(\pi) = \{v_1,\ldots,v_k\}$ for the set of seeds probed by $\pi$, $C(\pi)$ as shorthand for $C(S(\pi))$ for the set of vertices activated by $\pi$, and $\sigma(\pi) = \vert C(\pi)\vert$ for the influence of $\pi$.
We use $\mathbf{x}(\pi)=(x_v(\pi))_{v\in V}$ as the vector of \emph{seed probabilities}, where $x_v(\pi)=\Pr(v\in S(\pi))$.
For any $\mathbf{x}\in[0,1]^V$, we denote by $S_{\mathbf{x}}$ a random seed set drawn independently of $A$ by including each vertex $v\in V$ independently with probability $x_v$.
We denote the set of policies with budget $k$ by $\Pi_k$. We define the value of the best adaptive policy as
\begin{equation*}
    \Adaptive_k(V,\mathbf{p})=\max_{\pi\in\Pi_k}\E[\sigma(\pi)].
\end{equation*}
We omit the parameters $(V,\mathbf{p})$ from $\Static_k$ and $\Adaptive_k$ when the instance is clear from the context.
We will use that $\Adaptive_k\le k\Static_1$, which stems from adaptive submodularity~\cite{GolovinK11} and provides the elementary upper bound of $k$ on the adaptivity gap.

The \emph{adaptivity gap} of an instance $(V,\mathbf{p})$ with seed budget $k$ is given by the ratio \[\Adaptive_k(V,\mathbf{p})/\Static_k(V,\mathbf{p}),\] which is always at least $1$.
We discuss the problem for unweighted vertices.
More generally, each vertex $v$ may have a positive weight, and the influence of a seed set is the total weight of its cascade.
All our proofs extend immediately to weighted vertices, and we use unweighted vertices for simplicity of exposition. In general, integer-weighted instances can be exactly replaced by unweighted instances by increasing the number of vertices.\footnote{
    For a vertex $v$ with desired weight $w_v$, attach $w_v - 1$ new vertices to it, each live with probability $1$.
    The resulting group activates together, and seeding any of its vertices is equivalent to seeding $v$.}

\subsection{Our Results}

We establish three upper bounds on the adaptivity gap for the independent cascade model with full-adoption feedback.
For general undirected graphs, we obtain the first upper bound that is sublinear in the seed budget $k$ and independent of the graph size, improving on the elementary linear bound.
\begin{theorem}[Undirected graphs]
    \label{thm:undirected_sqrt_k}
    For every undirected instance of influence maximization,
    \begin{equation*}
        \Adaptive_k \leq 5\sqrt{k}\,\Static_k.
    \end{equation*}
\end{theorem}

We derive this bound by comparing the influence of each seed of an adaptive policy with the influence that this seed would add to a fixed set $S$ of seeds.
Since connections to $S$ are positively correlated, a seed may also activate vertices that $S$ reaches, and we control this overlap by splitting the seeds according to their probability of connecting to $S$.
For seeds that rarely connect to $S$, extending a percolation result of \citet{Gladkov24}, we show that a seed is unlikely to reach both a vertex and $S$ unless that vertex is itself likely to connect to $S$.
For seeds that connect to $S$ with probability at least $\beta$, at most one of them activates each vertex, so their total overlap is at most the influence of the set they form.
We bound this influence using a recently proved result conjectured by \citet{KozmaN24} in the context of percolation.
Combining both bounds over all thresholds $\beta$ yields the factor $\sqrt{k}$.
We give the proof of \cref{thm:undirected_sqrt_k} in \cref{sec:undirected}.
We then extend these results to any matroid constraint over the set of seeds and obtain a bound of $5\sqrt{r}$, where $r$ is the rank of the matroid; see \cref{cor:undirected_matroid}.

For general directed graphs, we improve the upper bound of $\lceil n^{1/3}\rceil$~\cite{DAngeloPV23}, obtaining the first sublogarithmic upper bound.

\begin{theorem}[Directed graphs]
\label{thm:directed_log_n}
    For every directed instance of influence maximization with $n$ vertices,
    \begin{equation*}
        \Adaptive_k \leq \mathcal{O}_{n\rightarrow\infty}\left(\frac{\log n}{\log\log n}\right)\Static_k.
    \end{equation*}
\end{theorem}

We obtain this result by studying an auxiliary batched policy that repeatedly seeds an optimal set of at most $k$ vertices in the remaining graph until no high-influence vertex remains.
We show that the probability that a vertex retains high influence decreases geometrically with the number of batches, so the policy needs only a sublogarithmic number of batches in expectation.
We then compare this policy with any adaptive policy by bounding each batch's expected additional influence by $\Static_k$ and using the absence of high-influence vertices at termination to control the additional influence obtainable from $k$ adaptive seeds.
We give the proof of \cref{thm:directed_log_n} in \cref{sec:directed}.

We also give an upper bound of $2$ for bi-directed forests, unifying several previous bounds for specific orientations of forests and giving the first constant upper bound for undirected forests.
A \emph{bi-directed forest} is obtained from an undirected forest by replacing each edge by two independent arcs in opposite directions with possibly different live probabilities.

\begin{theorem}[Bi-directed forests]
    \label{thm:bi_forest}
    For every bi-directed forest instance of influence maximization,
    \begin{equation*}
        \Adaptive_k \leq 2\,\Static_k.
    \end{equation*}
\end{theorem}

The proof builds on the depletion idea used for general directed graphs in~\cref{sec:directed}.
The forest structure lets us choose a small seed set whose cascade leaves only vertices with small expected influence.
We then bound the additional influence that any adaptive policy can obtain beyond this cascade, giving an adaptivity gap of at most $2$.
We give the proof of~\cref{thm:bi_forest} in~\cref{sec:forest}.

\cref{tab:results} summarizes our bounds alongside the best previously known upper bounds.

\begin{table}[H]
    \centering
    \begin{tabular}{llc}
        \toprule
        Graph Class & Previous Upper Bounds & This Work \\
        \midrule
        Undirected graphs & $\lceil n^{1/3}\rceil,\ \sqrt{\alpha}+\mathcal{O}(1)$~\cite{DAngeloPV23} & $5\sqrt{k}$ \\
        Directed graphs & $\lceil n^{1/3}\rceil$~\cite{DAngeloPV23} & $\mathcal{O}(\log n/\log\log n)$ \\
        In-arborescences & $\frac{2e^2}{e^2-1}$~\cite{DAngeloPV23} & $2$ \\
        Out-arborescences & $2$~\cite{ChenP19} & $2$ \\
        Bi-directed forests & $\lceil n^{1/3}\rceil,\ \sqrt{\alpha}+\mathcal{O}(1)$~\cite{DAngeloPV23} & $2$ \\
        \bottomrule
    \end{tabular}
    \caption{Adaptivity-gap upper bounds under full-adoption feedback. Here $\alpha$ is the sum of the degrees of vertices of degree greater than two in the underlying undirected graph.}
    \label{tab:results}
\end{table}

Finally, we improve the best known lower bound for influence maximization from $e/(e-1)\approx1.58$~\cite{ChenP19} to $1+\varphi\approx2.618$, where $\varphi$ is the golden ratio.
\begin{theorem}[Lower bound]
\label{thm:lower_bound}
Over directed instances of influence maximization,
    \[
        \sup_{V,\mathbf{p},k}\frac{\Adaptive_k(V,\mathbf{p})}{\Static_k(V,\mathbf{p})}\ge1+\varphi.
    \]
\end{theorem}
The previous lower bound was based on a reduction to submodular stochastic probing under matroid constraints with coverage functions, for which $e/(e-1)$ is tight. 
Our improvement is conceptually important, as it establishes that the nature of influence maximization is different from that of submodular stochastic probing. 
The proof is surprisingly technical and relies on a general reduction that shows that the worst-case adaptivity gap is equal to the worst-case gap between any adaptive policy and independent sampling from that policy's seed probabilities. 
The lower bound already holds for the class of directed acyclic graphs (DAGs).
We give the proof of \cref{thm:lower_bound} in \cref{sec:lb}.

We discuss the connections with previous work in the following subsection.
We hope that the techniques developed here will also prove useful for other problems in influence maximization and, more generally, for graph diffusion problems.

\subsection{Related Work}

\paragraph{Influence Maximization.}
\citet*{KempeKT03} formulated influence maximization as a discrete optimization problem for the independent cascade and the linear threshold models and established approximation guarantees through submodularity.
The connection between diffusion and submodularity extends beyond the independent cascade model: \citeauthor{MosselR10}~\cite{mossel2007submodularity,MosselR10} showed that monotone submodular local activation functions yield a monotone submodular expected influence function in the general threshold model.

A substantial body of work addresses the cost of evaluating and maximizing influence.
\citet*{ChenWW10} proved that computing influence exactly under the independent cascade model is $\#\mathrm{P}$-hard.
Efficient approximation algorithms can achieve a ratio of $(1-1/e-\varepsilon)$ without exact expected influence evaluation~\cite{BorgsBCL14, TangXS14}.
For undirected networks in the static model, \citet{KhannaL14} used properties of bond percolation to show an improved approximation ratio for the greedy algorithm.
Our analysis leverages several tools from percolation theory to study the adaptivity gap in undirected networks.
For adaptive policies, adaptive greedy achieves a $(1-1/e)$-approximation under full-adoption feedback~\cite{GolovinK11}, while its best known approximation ratio under myopic feedback is $1-1/\sqrt{e}$~\cite{DAngeloPV21}.

\paragraph{Adaptivity Gaps.}
The concept of adaptivity gap was introduced by \citet{DeanGV04} as a quantification of the gain of adaptivity in stochastic decision problems.
Adaptivity gaps have since been studied in stochastic submodular maximization and stochastic probing.
For independent item outcomes and a matroid constraint,~\citet{AsadpourN16} proved the tight $e/(e-1)$ bound and gave a polynomial-time stochastic continuous greedy algorithm that computes a static policy within a $(1-1/e-\varepsilon)$ factor of the adaptive optimum.
Constant adaptivity gaps for submodular objectives under prefix-closed probing constraints were established by \citet*{GuptaNS17}.
For monotone submodular objectives under these constraints, a tight bound of $2$ was proved by \citet*{BradacSZ19}.
In contrast, under XOS or subadditive valuations, the adaptivity gap is not constant in general~\cite{GuptaNS17, li2025adaptivity}.
Adaptivity gaps have been analyzed in diverse settings, such as stochastic orienteering~\cite{bansal2015adaptivity}, the stochastic traveling salesman problem~\cite{jiang2020stochastic}, and assortment optimization~\cite{el2026two}.
Bounding the adaptivity gap remains a major open problem in the context of influence maximization.

For influence maximization, under full-adoption feedback, \citet{ChenP19} proved upper bounds of $2e/(e-1)$ for in-arborescences and $2$ for out-arborescences, and a tight bound of $e/(e-1)$ for one-directional bipartite graphs.
They also proved a lower bound of $e/(e-1)$ even on directed paths.
\citet*{DAngeloPV23} improved the in-arborescence bound to $2e^2/(e^2-1)$ and established an upper bound of $\lceil n^{1/3}\rceil$ for general directed graphs.
They also obtained a $\sqrt{\alpha}+\mathcal{O}(1)$ bound, where $\alpha$ is the sum of degrees greater than two in the underlying undirected graph; opposite arcs may have different live probabilities.
Their results cover several restricted classes of bi-directed forests, for which our theorem improves the upper bound to $2$ simultaneously.

Under myopic feedback, \citet{PengC19} showed that the adaptivity gap lies between $e/(e-1)$ and $4$.
\citet*{DAngeloPV21} subsequently improved the upper bound to $2e/(e-1)$.

\paragraph{Percolation Inequalities.}
The conjecture of the absence of percolation at criticality was a central open problem in percolation theory concerning the existence of infinite connected components in random graphs.
Consider the lattice graph with vertex set $\Z^d$, where two vertices are adjacent if their $\ell_1$ distance is $1$.
Each edge is present independently with probability $p$.
There exists a critical probability $p_c$ such that, for $p < p_c$, all connected components are almost surely finite, whereas for $p > p_c$, an infinite component exists almost surely~\cite{Grimmett99}.
The conjecture asked whether all components remain finite almost surely at $p = p_c$.
Equivalently, it asks whether $\theta(p_c) = 0$, where $\theta(p)$ denotes the probability that the origin belongs to an infinite component.
The conjecture had been proved for $d = 2$~\cite{Harris60,Kesten80} and for $d \geq 19$~\cite{HaraS90}, with the latter result subsequently extended to $d \geq 11$~\cite{FitznerH17}.

\citet{KozmaN24} reduced the conjecture in all dimensions to connection inequalities for finite random graphs with arbitrary independent edge probabilities.
These inequalities were recently formalized in Lean and proved with the help of AI systems~\cite{Leder26,BouRabee26,OpenAI26}. 
Subsequently, \citet{BlancRenaudie26} and \citet{Gouere26} provided simpler, human-readable proofs.
We use the first inequality in \citet{KozmaN24} (stated as Conjecture 1 in that work) in our proof of the bound for undirected graphs.

We also use the three-point inequality of \citet[Thm.~6.2]{Gladkov24}, which bounds the probability that three vertices belong to the same connected component in terms of their three pairwise connection probabilities.\footnote{The same inequality with a better constant was also claimed by \citet{KhannaL14} and applied in the context of influence maximization. However, we encountered gaps in the original argument that we could not resolve.}
We extend this inequality to a set in place of one of the vertices; see \cref{lem:3PI}.

\section{An \texorpdfstring{$\mathcal{O}(\sqrt{k})$}{O(sqrt(k))} Upper Bound for Undirected Graphs}
\label{sec:undirected}

In this section, we prove the upper bound of \cref{thm:undirected_sqrt_k} for undirected instances.
An instance $(V, \mathbf{p})$ is \emph{undirected} if $p_{u,v} = p_{v,u}$ for every $u, v \in V$.
In the independent cascade model, this is equivalent to having undirected edges $\set{u, v}$ that are live independently with probability $p_{u,v}$.\footnote{For each pair $\set{u, v}$, only the arc from the endpoint activated first to the other endpoint can produce a new activation.
    Since both arcs have the same live probability, we may represent the pair by a single independent undirected edge.
    Even after observing earlier cascades, the arcs between vertices not yet activated remain independent with their original probabilities, so this replacement also applies to adaptive seed choices.
    Consequently, $\Static_k$ and $\Adaptive_k$ are the same in both representations.
}
We therefore index $\mathbf{p}$ by unordered pairs $\set{u, v} \subseteq V$.
In this representation, $A$ denotes the random set of live undirected edges.

For vertices $u, v \in V$, we write $u \conn v$ when they are connected by a path of live edges; equivalently, $v \in C(u)$, or $u \in C(v)$, or $C(u) = C(v)$, each of these conditions implying the others.
For a set $S \subseteq V$, we write $u \conn S$ if $u \in C(S)$, equivalently if $C(u) \subseteq C(S)$.
Throughout this section, we write $\mathbf{x}=\mathbf{x}(\pi)$ for the seed probabilities of the policy $\pi$ under consideration.
Our proof revolves around establishing the following lemma.

\begin{lemma}
    \label{lem:fixedset}
    Let $(V, \mathbf{p})$ be an undirected instance of influence maximization and $\pi$ be an adaptive policy with seed probabilities $\mathbf{x}$.
    Then, for every vertex $u \in V$ and every set $S \subseteq V$,
    \begin{equation*}
        \Pr(u \conn S(\pi)) \leq 2\sum_{v \in V} x_v\Pr(u \conn v,\,u \nconn S) + 4\sqrt{\sum_{v \in V} x_v}\Pr(u \conn S).
    \end{equation*}
\end{lemma}

Notice that, similarly to the function $S\mapsto\E[\sigma_V(S)]$, the function $f:S\mapsto\Pr(u\conn S)$ is monotone and submodular, and that $\Pr(u\conn v,\,u\nconn S)=f(S\cup\{v\})-f(S)$. Thus, it holds that
\begin{equation*}
    f^+(\mathbf{x})\le\min_{S\subseteq V}\left\{f(S)+\sum_{v\in V}x_v[f(S\cup\{v\})-f(S)]\right\},
\end{equation*}
where $f^+$ is the concave closure of $f$~\cite{Vondrak07}. When $\sum_{v \in V} x_v \leq k$, \cref{lem:fixedset} can be seen as an analog of this inequality up to a factor $\mathcal{O}(\sqrt{k})$, providing an upper bound on the performance of any adaptive policy $\pi$, which may both correlate seeds and choose them depending on the graph realization.

\begin{proof}[Proof of \cref{thm:undirected_sqrt_k}]
    Let $\pi$ be an adaptive policy with $\sum_{v \in V} x_v \leq k$.
    Let $S $ be a set of $k$ seeds maximizing $\E[\sigma(S)]$, so that $\E[\sigma(S)] = \Static_k$.
    Fix a vertex $v \in V$.
    Optimality of $S$ and submodularity of $\E[\sigma(\cdot)]$ then imply that $\E[\sigma(S\cup\{v\})-\sigma(S)] \le \E[\sigma(S)]/k$~\cite{NemhauserWF78}.
    By summing~\cref{lem:fixedset} over all vertices $u\in V$, we get, for $k\ge 4$,
    \begin{align*}
        \E[\sigma(\pi)]&\le2\sum_{v\in V}x_v\E[\sigma(S\cup\{v\})-\sigma(S)]+4\sqrt{\sum_{v\in V}x_v}\E[\sigma(S)]\\
        &\le2\E[\sigma(S)]+4\sqrt{k}\E[\sigma(S)]\le 5\sqrt{k}\Static_k.
    \end{align*}
    The second inequality stems from $\sum_vx_v\le k$. For $k\le 3$, the bound is elementary.
\end{proof}

Notably, this result requires the adaptive policy to satisfy the budget constraint only in expectation, while the static solution must satisfy it in every realization.
Our techniques allow us to generalize the budget constraint to matroid constraints; see \cref{cor:undirected_matroid}.

We now focus on proving Lemma~\ref{lem:fixedset}. Throughout this proof, we fix a static seed set $S\subseteq V$, an adaptive policy $\pi\in\Pi_k$, and a vertex $u\in V$.
The \emph{contribution} of a possible seed $v$ is the probability that $v$ is probed by $\pi$ and activates $u$, meaning that $u$ is connected to $v$ but is not connected to the seeds probed by $\pi$ before $v$.
The sum of these contributions over $v\in V$ is $\Pr(u \conn S(\pi))$.
We define the \emph{excess} of a seed $v$ as
\begin{equation*}
    e(v) \coloneqq \paren{\Pr(v \text{ activates } u) - 2x_v\Pr(u \conn v,\,u \nconn S)}_+,
\end{equation*}
where $y_+ = \max(y,0)$. The probability $\Pr(u \conn v,\,u \nconn S)$ is the marginal increase in the activation probability of $u$ when $v$ is added to $S$.
Thus, $e(v)$ measures how much $v$'s contribution under $\pi$ exceeds twice this marginal probability weighted by $x_v$.
We bound this excess using the probability that $v$ connects to both $u$ and $S$.
Since $\Pr(u \conn S(\pi)) \leq 2\sum_{v \in V} x_v\Pr(u \conn v,\,u \nconn S) + \sum_{v \in V} e(v)$, it suffices to show that $\sum_{v \in V} e(v) \leq 4\sqrt{\sum_{v \in V} x_v}\Pr(u \conn S)$.
The excess is related to the probability $\kappa(v) \coloneqq \Pr(v \conn S)$ that $v$ connects to $S$.
We bound the excess individually for seeds with small $\kappa(v)$, which we call \emph{weakly connected}, and collectively for seeds with large $\kappa(v)$, which we call \emph{strongly connected}.

\medskip
\noindent
\textbf{Weakly Connected Seeds.} For a seed $v$ to have positive excess, it must be able to reach both $u$ and $S$.
At the time of choosing $v$, the edges between the still inactive vertices follow their original distribution.
Thus, whenever $v$ is selected, its probability of activating $u$ is at most $\Pr(u \conn v)$, and since $v$ is selected with probability $x_v$, we have that $\Pr(v \text{ activates } u) \leq x_v\Pr(u \conn v)$.
Since $\Pr(u \conn v) = \Pr(u \conn v,\,v \conn S) + \Pr(u \conn v,\,u \nconn S)$, we obtain that
\begin{equation}
    \label{eqn:excess_weak_bound}
    e(v) \leq x_v\paren{\Pr(u \conn v,\,v \conn S) - \Pr(u \conn v,\,u \nconn S)}_+.
\end{equation}
To bound the first term of the expression above, we use a slight generalization of the three-point inequality of \citet{Gladkov24}.

\begin{lemma}[Three-point inequality with sets]
    \label{lem:3PI}
    For all vertices $u, v \in V$ and every set $S \subseteq V$,
    \begin{equation*}
        \Pr(u \conn v,\ v \conn S)^2 \leq 8\Pr(u \conn v)\,\Pr(v \conn S)\,\Pr(u \conn S).
    \end{equation*}
\end{lemma}

\begin{proof}[Proof of \cref{lem:3PI}]
If $S=\emptyset$ or $\Pr(u \conn v)=0$, the claim is immediate. Otherwise, write $\Pr^\star$ for probabilities in the instance obtained by contracting $S$ into a single vertex $s^\star$. Identifying each vertex with its image under contraction, we have $\Pr^\star(w \conn s^\star)=\Pr(w \conn S)$ for every $w\in V$.

Contraction creates a $u$--$v$ connection precisely when $u$ and $v$ lie in distinct components that both meet $S$. These are also the realizations that contraction adds to the event $\{u \conn v,\,v \conn S\}$. Thus, contraction can only increase both $\Pr(u \conn v,\,v \conn S)$ and $\Pr(v \conn S \mid u \conn v)$, since it adds the same probability to the numerator and denominator of the latter. Taking the product of these two probabilities and applying the three-point inequality of \citet[Thm.~6.2]{Gladkov24} in the contracted graph gives
\begin{equation*}
    \frac{\Pr(u \conn v,\,v \conn S)^2}{\Pr(u \conn v)} \leq \frac{\Pr^\star(u \conn s^\star,\,v \conn s^\star)^2}{\Pr^\star(u \conn v)} \leq 8\Pr^\star(u \conn s^\star)\Pr^\star(v \conn s^\star) = 8\Pr(u \conn S)\Pr(v \conn S).
\end{equation*}
Multiplying by $\Pr(u \conn v)$ proves the claim.
\end{proof}

Thus, by \cref{lem:3PI},
\begin{equation*}
    \Pr(u \conn v,\,v \conn S) \leq \sqrt{8\Pr(u \conn v)\kappa(v)\kappa(u)} \leq \frac{\Pr(u \conn v)}{2} + 4\kappa(v)\kappa(u),
\end{equation*}
where the second inequality follows from the AM--GM inequality.
Rearranging and applying \eqref{eqn:excess_weak_bound} gives
\begin{equation}
    \label{eq:weak-excess}
    e(v) \leq 8x_v\kappa(v)\kappa(u) \qquad \text{for all } v \in V.
\end{equation}

\medskip
\noindent
\textbf{Strongly Connected Seeds.} For strongly connected seeds, bounding the excess of each seed separately is too loose.
These individual bounds may count several seeds that connect to $u$, although at most one activates it in an execution of $\pi$.
We therefore bound the total contribution of the strongly connected seeds at once.
Since $u$ is activated by at most one seed, the probability that this seed belongs to a set $T \subseteq V$ is at most $\Pr(u \conn T)$.
Let $T_\beta \coloneqq \setst{v \in V}{\kappa(v) \geq \beta}$ be the set of vertices that connect to $S$ with probability at least $\beta$.
Since $T_\beta$ does not depend on the realization, we obtain that
\begin{equation*}
    \sum_{v \in T_\beta} e(v) \leq \sum_{v \in T_\beta}\Pr(v \text{ activates } u) \leq \Pr(u \conn T_\beta).
\end{equation*}
It remains to compare $\Pr(u \conn T_\beta)$ with $\Pr(u \conn S)$. To do this, we leverage the Kozma--Nitzan inequality~\cite{KozmaN24,Gouere26}, formulated in the context of percolation, which we state as follows for all vertices $u, v\in V$ and every nonempty set $T\subseteq V$:
\begin{equation}
    \label{eqn:kz_conj_1}
    \Pr(u \conn v) \geq \Pr(u \conn T)\min_{t \in T}\Pr(t \conn v).
\end{equation}
If $T_\beta \ne \emptyset$, write $\Pr^\star$ for probabilities in the instance obtained by contracting $S$ into a single vertex $s^\star$, as in the proof of \cref{lem:3PI}.
Contraction preserves connections to $S$, so $\Pr^\star(w \conn s^\star)=\Pr(w \conn S)$ for every $w\in V$, while it can only increase the probability of connecting $u$ to $T_\beta$.
Applying \eqref{eqn:kz_conj_1} to the contracted graph therefore gives
\begin{equation*}
    \Pr(u \conn S) = \Pr^\star(u \conn s^\star) \geq \Pr^\star(u \conn T_\beta)\min_{t \in T_\beta}\Pr^\star(t \conn s^\star) \geq \beta\Pr^\star(u \conn T_\beta) \geq \beta\Pr(u \conn T_\beta).
\end{equation*}
Combining both inequalities gives the following bound for strongly connected seeds:
\begin{equation}
    \label{eq:strong-excess}
    \sum_{v \in V:\,\kappa(v) \geq \beta} e(v) \leq \kappa(u)/\beta \qquad \text{for all } 0 < \beta \leq 1.
\end{equation}

\medskip
\noindent
\textbf{Merging the Bounds.} Finally, we combine the bounds into a proof of \cref{lem:fixedset}.

\begin{proof}[Proof of \cref{lem:fixedset}]
It suffices to show that $\sum_{v \in V} e(v) \leq 4\sqrt{\sum_{v \in V} x_v}\,\kappa(u)$.
By \eqref{eq:weak-excess}, $e(v) = 0$ whenever $\kappa(v) = 0$, and
\begin{equation*}
    \sum_{v \in V:\,\kappa(v) > 0} \frac{e(v)}{\kappa(v)} \leq 8\kappa(u)\sum_{v \in V} x_v.
\end{equation*}
On the other hand, applying \eqref{eq:strong-excess} with $\beta = \kappa(v)$ gives $\sum_{t \in V:\,\kappa(t) \geq \kappa(v)} e(t) \leq \kappa(u)/\kappa(v)$ for every $v$ with $\kappa(v) > 0$.
Combining both bounds gives
\begin{equation*}
    \paren*{\sum_{v \in V} e(v)}^2 \leq 2\sum_{v \in V:\,\kappa(v) > 0} e(v)\sum_{t \in V:\,\kappa(t) \geq \kappa(v)} e(t) \leq 2\kappa(u)\sum_{v \in V:\,\kappa(v) > 0}\frac{e(v)}{\kappa(v)} \leq 16\kappa(u)^2\sum_{v \in V} x_v.
\end{equation*}
Taking square roots on both sides concludes the proof.
\end{proof}

\medskip
\noindent
\textbf{Matroid Constraints.} We show that~\cref{thm:undirected_sqrt_k} extends beyond cardinality constraints to matroid constraints.
We first derive the following consequence of \cref{lem:fixedset}, which compares an adaptive policy with independent sampling according to its seed probabilities.

\begin{proposition}
    \label{prop:undirected_rounding}
    Let $(V, \mathbf{p})$ be an undirected instance of influence maximization and $\pi$ be an adaptive policy with seed probabilities $\mathbf{x}$.
    Then,
    \begin{equation*}
        \Pr(u \conn S_{\mathbf{x}}) \geq \frac{\Pr(u \conn S(\pi))}{2+4\sqrt{\sum_{v \in V} x_v}} \qquad \text{for all } u \in V.
    \end{equation*}
\end{proposition}
\begin{proof}
    Fix a vertex $u \in V$.
    Order the vertices of $V$ as $v_1,\dots,v_n$ and write $S_i = S_{\mathbf{x}} \cap \set{v_1,\dots,v_i}$, with $S_0 = \emptyset$.
    Adding $v_i$ activates $u$ exactly when $v_i$ is sampled, $u \conn v_i$, and $u \nconn S_{i-1}$.
    By independence of the sampling and since $S_{i-1} \subseteq S_{\mathbf{x}}$,
    \begin{equation*}
        x_{v_i}\Pr(u \conn v_i,\,u \nconn S_{\mathbf{x}}) \leq x_{v_i}\Pr(u \conn v_i,\,u \nconn S_{i-1}) = \Pr(u \conn S_i)-\Pr(u \conn S_{i-1}).
    \end{equation*}
    Summing over $i$, we obtain that $\sum_{v \in V} x_v\Pr(u \conn v,\,u \nconn S_{\mathbf{x}}) \leq \Pr(u \conn S_{\mathbf{x}})$.
    Since $S_{\mathbf{x}}$ is independent of $A$, averaging \cref{lem:fixedset} over $S_{\mathbf{x}}$ concludes the proof.
\end{proof}

A \emph{matroid} $\calM = (V, \calI)$ is a pair where $V$ is a finite set of elements, called the \emph{ground set}, and $\calI \subseteq 2^V$ is a family of subsets whose members are called \emph{independent sets}. The family $\calI$ satisfies three properties: (i) $\varnothing \in \calI$; (ii) if $I_1 \in \calI$, then $I_2 \in \calI$ for all $I_2 \subseteq I_1$; (iii) if $I_1, I_2 \in \calI$ with $\card{I_1} > \card{I_2}$, then there exists $w \in I_1 \setminus I_2$ such that $I_2 \cup \set{w} \in \calI$.

The \emph{rank} of $\calM$ is $r=\max_{I\in\calI}\card{I}$.
Let $\Static_{\calM}$ and $\Adaptive_{\calM}$ denote the optimal expected influence of static and adaptive policies, respectively, whose seed sets are independent in $\calM$ in every execution.
Swap rounding~\cite{ChekuriVZ10} rounds the vector $\mathbf{x}(\pi)$ of a feasible adaptive policy $\pi$ to a random independent set with expected influence at least that of $S_{\mathbf{x}}$.
Since seed probabilities sum to at most the rank $r$, \cref{prop:undirected_rounding} gives the factor $2+4\sqrt{r}\le 5\sqrt{r}$ for $r\ge 4$, extending \cref{thm:undirected_sqrt_k} to matroid constraints.
For $r\le 3$, the bound is elementary. 
Which gives the following result.

\begin{cor}[Matroid constraints]
    \label{cor:undirected_matroid}
    For every undirected instance of influence maximization and every matroid $\calM$ on the vertex set with rank $r$,
    \begin{equation*}
        \Adaptive_{\calM} \leq 5\sqrt{r}\,\Static_{\calM}.
    \end{equation*}
\end{cor}

\section{A Sublogarithmic Upper Bound for Directed Graphs}
\label{sec:directed}
In this section, we prove the upper bound of~\cref{thm:directed_log_n} for directed instances.
Our proof revolves around analyzing an auxiliary policy that repeatedly seeds an optimal batch of at most $k$ inactive vertices until no high-influence vertex remains.
For a suitable high-influence threshold, we show that the expected number of batches before depletion is sublogarithmic.

Let $(V,\mathbf{p})$ be an instance. For $U\subseteq V$, we write $\restr{\mathbf{p}}{U}$ and $A[U]$ for the restrictions of $\mathbf{p}$ and $A$ to arcs with both endpoints in $U$.
The instance restricted to $U$ is $(U,\restr{\mathbf{p}}{U})$.
For $S\subseteq V$, we interpret $C_U(S)$ as the vertices reachable from $S\cap U$ in $A[U]$; in particular, $\sigma_U(v) = 0$ for $v\notin U$.
For fixed $U$, the arcs of $A[U]$ are independent with their original probabilities.
This remains true conditional on any cascade history with inactive set $U$, since exploration only reveals arcs from activated vertices.

Fix a parameter $b > 1$ and let $H(U)=\{v\in U:\E[\sigma_U(v)]\ge b\Static_k(V,\mathbf{p})\,/\,k\}$ be the set of \emph{high-influence} vertices of $U$.
Let $T$ be the random number of batches used by~\cref{alg:batched_policy}. We prove the following bound.

\begin{lemma}\label{lemma:expected_depletion} For every $b>1$,
    \[
        \E[T]\le\left\lceil\frac{\log n}{\log b}\right\rceil + \frac{b}{b-1}.
    \]
\end{lemma}

\begin{algorithm}[!t]
    \caption{Batched depletion policy}
    \label{alg:batched_policy}
    \setlength{\topsep}{0pt}
    \setlength{\partopsep}{0pt}
    \begin{tabbing}
        \hspace{1.5em}\=\kill
        \textbf{Input:} Instance $(V,\mathbf{p})$, budget $k$, parameter $b$\\
        $U_1 \gets V$\\
        $t \gets 1$\\
        \textbf{while} $H(U_t)\neq\varnothing$ \textbf{do}\\
        \> Choose $S_t\subseteq U_t$ with $|S_t|\le k$ such that $\E[\sigma_{U_t}(S_t)\mid U_t]=\Static_k(U_t,\restr{\mathbf{p}}{U_t})$\\
        \> Seed all vertices in $S_t$ and observe $C_{U_t}(S_t)$\\
        \> $U_{t+1} \gets U_t\setminus C_{U_t}(S_t)$\\
        \> $t\gets t+1$\\
        \textbf{end while}
    \end{tabbing}
\end{algorithm}

For constant $b$, this gives $\mathcal{O}(\log n)$ expected batches; choosing $b$ as a function of $n$ will give the sublogarithmic bound.
Following the definitions of \cref{alg:batched_policy}, for $t>T$, set $S_t=\emptyset$ and $U_t=U_{T+1}$.

\begin{proof}[Proof of \cref{lemma:expected_depletion}]
Deleting vertices cannot increase the expected influence of any remaining vertex, so $U_t\subseteq U_{t-1}$ implies $H(U_t)\subseteq H(U_{t-1})$.
Fix $t\ge 2$ and suppose $H(U_{t-1})\ne\emptyset$, so batch $t-1$ is executed.
Furthermore, we have the uniform bound
\[
    \E[\sigma_{U_{t-1}}(S_{t-1})\mid U_{t-1}]=\Static_k(U_{t-1},\restr{\mathbf{p}}{U_{t-1}})\le\Static_k(V,\mathbf{p}).
\]
For every fixed $v\in V$, we have $\sigma_{U_t}(v)=\sigma_{U_{t-1}}(S_{t-1}\cup\{v\})-\sigma_{U_{t-1}}(S_{t-1})$.
Thus, by submodularity of $\sigma_{U_{t-1}}$ and optimality of $S_{t-1}$, we have that $\E[\sigma_{U_t}(v)\mid U_{t-1}]\le\Static_k(V, \mathbf{p})\,/\,k$~\cite{NemhauserWF78}.
By the tower property, it follows that
\begin{align*}
    \E[\E[\sigma_{U_t}(v)\mid U_t]\mid U_{t-1}]\le \frac{\Static_k}{k}.
\end{align*}
Markov's inequality and $H(U_t)\subseteq H(U_{t-1})$ therefore give
\begin{align*}
    \Pr(v\in H(U_t)\mid U_{t-1})\le\frac{1}{b}\Iverson{v\in H(U_{t-1})},
\end{align*}
where $\Iverson{\calE}$ denotes the \emph{indicator} of the event $\calE$.
Taking expectations and iterating from $U_1=V$ gives, for every $t\ge 1$,
\begin{align*}
    \Pr(v\in H(U_t))\le\frac{1}{b^{t-1}}.
\end{align*}
Since $\{T>t\}=\{H(U_{t+1})\ne\emptyset\}$, a union bound gives, for every integer $t\ge 0$,
\begin{align*}
    \Pr(T > t)&\le\min\left(1,\sum_{v\in V}\Pr(v\in H(U_{t+1}))\right)\le\min\left(1,\frac{n}{b^t}\right).
\end{align*}
Finally, summing the tail bound gives
\begin{equation*}
    \E[T] =\sum_{t\ge 0}\Pr(T>t)
    \le\sum_{t\ge 0}\min\left(1,\frac{n}{b^t}\right)
    \le \left\lceil\frac{\log n}{\log b}\right\rceil + \frac{b}{b-1}. \qedhere
\end{equation*}
\end{proof}

We can now prove \cref{thm:directed_log_n}.

\begin{proof}[Proof of \cref{thm:directed_log_n}]
    Let $D$ be the set of vertices activated by~\cref{alg:batched_policy}.
    For every $t\le T$, conditional on $U_t$, the expected additional influence of batch $t$ is $\Static_k(U_t,\restr{\mathbf{p}}{U_t})\le\Static_k(V, \mathbf{p})$, since the remaining arcs are independent with their original probabilities.
    Since $\{T\ge t\}=\{H(U_t)\neq\emptyset\}$, summing over the batches gives
    \begin{equation*}
        \E[|D|] =\sum_{t=1}^\infty\E[\Iverson{T\ge t}\sigma_{U_t}(S_t)]
        \leq \sum_{t=1}^\infty\Pr(T\ge t)\Static_k(V, \mathbf{p}) = \E[T]\Static_k(V, \mathbf{p}).
    \end{equation*}
The inequality follows by conditioning on $U_t$ in each term.
Take an adaptive policy $\pi\in\Pi_k$. Run \cref{alg:batched_policy}.
Then, replay $\pi$ in the same realization, using its original feedback $C(v_1),\ldots,C(v_{i-1})$ to choose $v_i$ and ignoring the depletion history.
We have that $C(\pi)\subseteq D \cup(C(\pi)\cap U_{T+1})$, and so
\begin{align*}
    \E[\sigma(\pi)]
    &\le \E[|D|]+\E\left[\E[|C(\pi)\cap U_{T+1}|\mid U_{T+1}]\right]\le \E[|D|]+\E[\Adaptive_k(U_{T+1},\restr{\mathbf{p}}{U_{T+1}})]\\
    &\le \E[|D|]+k\,\E[\Static_1(U_{T+1},\restr{\mathbf{p}}{U_{T+1}})]\le \E[T]\Static_k+k\,\frac{b\Static_k}{k}\\
    &=\left(b+\E[T]\right)\Static_k.
\end{align*}
The second inequality holds because no arc from $V\setminus U_{T+1}$ to $U_{T+1}$ is live, while, conditional on $U_{T+1}$, the arcs of $A[U_{T+1}]$ keep their original distribution. Thus, $C(\pi)\cap U_{T+1}$ is reached only through arcs of $A[U_{T+1}]$, and the remaining feedback of $\pi$ depends only on arcs independent of $A[U_{T+1}]$, which act as external randomness. Hence, $\pi$ induces an adaptive policy with budget $k$ on $(U_{T+1},\restr{\mathbf{p}}{U_{T+1}})$.

Choosing $b=\sqrt{\log n}$ and applying~\cref{lemma:expected_depletion} gives
\begin{align*}
    \E[\sigma(\pi)]&\le\left(\sqrt{\log n}+\left\lceil\frac{2\log n}{\log\log n}\right\rceil+\frac{\sqrt{\log n}}{\sqrt{\log n} -1}\right)\Static_k
    =\mathcal{O}_{n\rightarrow\infty}\left(\frac{\log n}{\log\log n}\right)\Static_k.
\end{align*}
Taking $\pi$ to be an optimal adaptive policy concludes the proof.
\end{proof}

\section{An Upper Bound of 2 for Bi-Directed Forests}
\label{sec:forest}

Formally, an instance $(V, \mathbf{p})$ is a bi-directed forest if the undirected graph on $V$ with an edge $\set{u, v}$ for each pair of distinct vertices with $p_{u,v} > 0$ or $p_{v,u} > 0$ is a forest.
Opposite arcs are live independently and may have different probabilities.
This includes every orientation of a forest (such as in- or out-arborescences) and, by the equivalence in \cref{sec:undirected}, undirected forests.
We show that the adaptivity gap of every such instance is at most $2$.
We use a standard result on packing and covering subtrees; see~\citet{GyarfasL70}. We include a short proof for completeness.

\begin{lemma}
    \label{lem:tree_hitting_set}
    Let $\calT = \set{T_1, \dots, T_m}$ be a family of nonempty subtrees of a forest $F$, and let $t \geq 1$ be an integer.
    Then either $\calT$ contains $t$ pairwise vertex-disjoint members, or there exists a set of at most $t - 1$ vertices of $F$ intersecting every member of $\calT$.
\end{lemma}
\begin{proof}
    We root each component of $F$ and begin with an empty family $\calL$ of subtrees and an empty set $L$ of vertices.
    While some member of $\calT$ is disjoint from $L$, we choose such a member $T$ whose vertex $v$ closest to the root is as deep as possible.
    Every member disjoint from $L$ that intersects $T$ contains $v$; otherwise, its vertex closest to the root would lie strictly below $v$, contradicting our choice of $T$.
    We add $T$ to $\calL$ and $v$ to $L$.

    Thus, the members of $\calL$ are pairwise vertex-disjoint, and upon termination $L$ intersects every member of $\calT$.
    Each step adds one new element to each set, so $\card{\calL} = \card{L}$.
    If $\card{\calL} \geq t$, the first alternative holds.
    Otherwise, $\card{L} = \card{\calL} \leq t - 1$, giving the second alternative.
\end{proof}

As in~\cref{sec:directed}, we first deplete vertices with large expected residual influence and then bound the additional influence of $k$ adaptive selections.
For forests, \cref{lem:tree_hitting_set} provides a fixed set $L$ of at most $k-1$ seeds whose removal leaves an instance in which any single seed has expected influence at most $\Static_k/k$.
Seeding $L$ contributes at most $\Static_k$ in expectation, and so do the $k$ further adaptive selections in total, giving the factor $2$.

\begin{proof}[Proof of \cref{thm:bi_forest}]
    Let $(V,\mathbf{p})$ be a bi-directed forest instance and let $F$ be its underlying undirected forest.
    As in~\cref{sec:directed}, the instance restricted to $U\subseteq V$ is $(U,\restr{\mathbf{p}}{U})$.
    Identify each subtree of $F$ with its vertex set, and let $\calT$ be the family of nonempty subtrees $T$ satisfying $\Static_1(T,\restr{\mathbf{p}}{T}) > \Static_k(V,\mathbf{p})/k$.
    There cannot be $k$ pairwise vertex-disjoint members of $\calT$.
    Otherwise, choosing an optimal seed in each would give expected influence greater than $\Static_k$, since their cascades (each limited to its own subtree) are disjoint.

    Applying \cref{lem:tree_hitting_set} to $F$ gives a set $L$ of at most $k - 1$ vertices intersecting every member of $\calT$.
    Every component of $F \setminus L$ is a subtree disjoint from $L$ and therefore does not belong to $\calT$.
    Since every directed live path avoiding $L$ stays within one such component, $\Static_1\paren{V \setminus L,\restr{\mathbf{p}}{V \setminus L}} \leq \Static_k(V,\mathbf{p})/k$.

    Fix an adaptive policy $\pi \in \Pi_k$ and compare its activations with $C(L)$ in the same realization.
    The vertices activated by $\pi$ inside $C(L)$ contribute at most $\sigma(L)$.
    At each step, a directed live path from the seed selected by $\pi$ to a newly activated vertex outside $C(L)$ avoids both $L$ and all vertices activated earlier by $\pi$.
    Consequently, each selection contributes at most $\Static_1\paren{V \setminus L,\restr{\mathbf{p}}{V \setminus L}}$ expected new activations outside $C(L)$.
    Since $\card{L} \leq k - 1$, summing over the $k$ selections gives
    \begin{equation*}
        \E[\sigma(\pi)] \leq \E[\sigma(L)] + k\,\Static_1\paren{V \setminus L,\restr{\mathbf{p}}{V \setminus L}}
        \leq \Static_{k - 1} + k \frac{\Static_k}{k}
        \leq 2\Static_k. \qedhere
    \end{equation*}
\end{proof}

\section{A Lower Bound of \texorpdfstring{$1+\varphi$}{1+φ}}
\label{sec:lb}

In this section, we establish the lower bound on the adaptivity gap as stated in \cref{thm:lower_bound}. 
We use $\varphi\approx1.618$ to denote the golden ratio, which is the largest solution of the equation $\varphi+1=\varphi^2$.

To simplify the exposition of the contents of this section, we allow vertices to have nonnegative weights. By scaling and rounding, and as remarked in \cref{sec:problem_definition}, these can be approximated with unit weights to arbitrary precision.
For a weight vector $\mathbf{w}\in\mathbb{R}_{\geq0}^V$, the influence $\sigma(S)$ is the total weight of the vertices in $C(S)$, and we define $\Static_k(V,\mathbf{p},\mathbf{w})$ and $\Adaptive_k(V,\mathbf{p},\mathbf{w})$ accordingly.

In \cref{sec:sample_gap}, we show that the worst-case adaptivity gap is equal to the supremum, over instances and adaptive policies $\pi$, of the \emph{sampling gap} $\E[\sigma(\pi)]/\E[\sigma(S_{\mathbf{x}(\pi)})]$. 
This equivalence is similar in spirit to the connection between adaptivity gaps and correlation gaps for monotone submodular functions~\cite{Vondrak07,agrawal2012price,AsadpourN16}, but its proof requires a new construction specific to influence maximization, based on connecting copies of the original instance through a large number of source vertices.
In \cref{sec:tree_phi}, we then construct a weighted tree instance with a sampling gap of $1+\varphi$.
This immediately implies the existence of instances with adaptivity gap $1+\varphi$, proving \cref{thm:lower_bound}.

\subsection{Adaptivity Gap Through Sampling Gap}
\label{sec:sample_gap}
We first show that the worst-case adaptivity gap matches the worst-case sampling gap.

\begin{lemma}
    \label{lem:sampling_gap}
    For influence maximization with nonnegative vertex weights,
    \begin{align*}
        \sup_{V,\mathbf{p},\mathbf{w},k}\frac{\Adaptive_k(V,\mathbf{p},\mathbf{w})}{\Static_k(V,\mathbf{p},\mathbf{w})}=\sup_{V,\mathbf{p},\mathbf{w},k}\,\sup_{\pi\in\Pi_k}\frac{\E[\sigma(\pi)]}{\E[\sigma(S_{\mathbf{x}(\pi)})]}.
    \end{align*}
\end{lemma}

\begin{proof}

    We begin by proving that the adaptivity gap is at most the sampling gap. 
    Fix an instance and let $\pi\in\Pi_k$ be an optimal adaptive policy.
    Since $S\mapsto\E[\sigma(S)]$ is monotone and submodular, swap rounding~\cite{ChekuriVZ10} applied to $\mathbf{x}(\pi)$ gives a random set $\widehat S$ of cardinality at most $k$ such that $\E[\sigma(\widehat S)]\geq\E[\sigma(S_{\mathbf{x}(\pi)})]$.
    Hence, some deterministic set $S$ of cardinality at most $k$ satisfies $\E[\sigma(S)]\geq\E[\sigma(S_{\mathbf{x}(\pi)})]$, and therefore
    \begin{equation*}
        \frac{\Adaptive_k}{\Static_k} \leq \frac{\E[\sigma(\pi)]}{\E[\sigma(S_{\mathbf{x}(\pi)})]}.
    \end{equation*}

    We now prove the reverse inequality.
    Fix an instance $G=(V, \mathbf{p}, \mathbf{w})$ and a policy $\pi\in\Pi_k$. Write $\mathbf{x}=\mathbf{x}(\pi)$ for the seed probabilities of $\pi$.
    For every integer $m \geq 1$, we construct an instance $H$ with seed budget $m$ such that
    \begin{align*}
        \Adaptive_m(H) &\geq m^2\,\E[\sigma_G(\pi)]-\mathcal{O}\paren{m^{3/2}},\\
        \Static_m(H) &\leq m^2\,\E[\sigma_G(S_{\mathbf{x}})]+\mathcal{O}\paren{m},
    \end{align*}
    where $\sigma_G$ denotes the influence in $G$ and $\mathcal{O}$ is taken as $m\to\infty$ with $G$ fixed.
    The construction simulates many independent runs of $\pi$ in parallel, one in each of $m^2$ copies of $G$, so that an adaptive policy $\pi_H$ on $H$ obtains nearly $\E[\sigma_G(\pi)]$ per copy.

    \medskip
    \noindent
    \paragraph{Instance.}
    The instance $H$ consists of $m^2$ disjoint copies $G_1,\dots,G_{m^2}$ of $G$ and a set $U$ of sources.
    Each copy keeps the vertex weights and arc probabilities of $G$, and we write $(i,v)$ for the copy of $v\in V$ in $G_i$.
    For every $t\in[m]$ and every set $J\subseteq[m^2]\times V$ that contains at most one vertex of each $G_i$ and at most $m$ copies of each $v\in V$, the set $U$ contains a source $u_{t,J}$ of weight zero, with an arc to each $(i,v)\in J$ that is live with probability $x_v$.
    All other arcs incident to $U$, and all arcs between different copies, have live probability zero.
    In particular, since the sources have no incoming arcs, $H$ is acyclic whenever $G$ is acyclic.

    \medskip
    \noindent
    \paragraph{Adaptive Policy.}
    We build an adaptive policy $\pi_H$ on $H$ with budget $m$ that simulates $m^2$ independent runs of $\pi$ in parallel, one in each copy.
    The policy $\pi_H$ seeds only sources, and seeding a source whose arc to $(i,v)$ is live plays the role of seeding $v$ in the run of $\pi$ in $G_i$.

    Formally, for each $i\in[m^2]$, the choices made by $\pi$ in $G_i$ use independent randomness across copies and depend only on the sequence of vertices $v\in V$ already seeded together with the resulting cascades.
    The run is \emph{finished} once $\pi$ selects no further vertex in it; until then, its \emph{current vertex} $a_i\in V$ is the vertex that $\pi$ seeds next, which changes only when an attempt at $a_i$ succeeds.
    In round $t=1, \dots, m$, a copy $G_i$ is \emph{active} if (1) $i \leq tm$, (2) its run is not finished, and (3) $G_i$ has not been abandoned in an earlier round.
    For each $v\in V$, if more than $m$ active copies have $a_i=v$, the policy $\pi_H$ abandons all but $m$ of them.
    It then seeds the source $u_{t,J}\in U$, where $J$ is the set of vertices $(i,a_i)$ over the remaining active copies, which makes an attempt at $a_i$ in each of these copies.
    The only arc from $u_{t,J}$ into $G_i$ ends at $(i,a_i)$.
    If this arc is live, the attempt succeeds: the cascade of $u_{t,J}$ in $G_i$ is the cascade of $(i,a_i)$, and the run in $G_i$ seeds $a_i$ and observes this cascade.
    Otherwise, the attempt fails and the run in $G_i$ does not progress in this round.
    The policy stops after round $m$, having seeded $m$ sources, so $\pi_H\in\Pi_m$.
    
    For the analysis, consider an unlimited run of $\pi$, which is never abandoned, continues after round $m$, and in each round after its start, called a local round, makes an attempt at its current vertex $v$ that succeeds independently with probability $x_v$.
    A failed attempt reveals no arc of the copy, so an unlimited run observes cascades with the same distribution as $\pi$ on $G$, and the expected weight it activates is $\E[\sigma_G(\pi)]$.
    We couple the run in each $G_i$ with an unlimited run in $G_i$ that starts in the same round, so that the two runs agree until $G_i$ is abandoned or round $m$ ends.
    This coupling is possible since each round uses a new source, so each attempt succeeds independently with probability $x_{a_i}$; we take the unlimited runs to be independent across copies.
    
    Let $T_v$ count the attempts of an unlimited run at $v$.
    The run selects $v$ with probability $x_v$, and at most once.
    If $x_v>0$, then conditional on this event, $T_v$ is geometric with mean $1/x_v$, so $\E[T_v]\le 1$.
    Let $N_{t,v}$ be the number of unlimited runs that make an attempt at $v$ in round $t$.
    Since $m$ runs start in each round,
    \begin{equation*}
        \E[N_{t,v}]
        =m\sum_{s=1}^{t}
          \Pr(\text{an unlimited run makes an attempt at $v$ in local round $s$})
        \le m\,\E[T_v]
        \le m.
    \end{equation*}
    Moreover, $N_{t,v}$ is a sum of independent indicators, so $\Var(N_{t,v})\le \E[N_{t,v}]\le m$, and therefore
    \begin{equation*}
        \E[(N_{t,v}-m)_+] \le \E\bigl[|N_{t,v}-\E[N_{t,v}]|\bigr] \le \sqrt{\Var(N_{t,v})} \le \sqrt{m}.
    \end{equation*}
    An active copy $G_i$ with $a_i=v$ in round $t$ has not been abandoned, so its run agrees with its unlimited run, which then makes an attempt at $v$ in round $t$.
    Hence, at most $(N_{t,v}-m)_+$ copies with $a_i=v$ are abandoned in round $t$, and the expected number of abandoned copies over all vertices and rounds is at most $nm\sqrt{m}=nm^{3/2}$.
    After abandonment in round $m$, at most $m$ active copies have $a_i=v$, for each $v\in V$, so at most $nm$ runs are unfinished when $\pi_H$ stops.
    
    If the run in $G_i$ is neither abandoned nor unfinished, it agrees with its unlimited run, and $\pi_H$ activates in $G_i$ the weight that this unlimited run activates; otherwise, $\pi_H$ activates at least this weight minus $w(V) = \sum_{v\in V}w_v$.
    Summing over the $m^2$ copies and writing $\sigma_H$ for the influence in $H$, we obtain
    \begin{equation*}
        \Adaptive_m(H) \geq \E[\sigma_H(\pi_H)] \geq m^2\,\E[\sigma_G(\pi)]-w(V)\,n\bigl(m^{3/2}+m\bigr).
    \end{equation*}

    \paragraph{Static Policy.}
    Fix a set $S$ of at most $m$ seeds in $H$.
    For $i\in[m^2]$ and $v\in V$, let $h_{i,v}$ be the number of sources in $S$ with an arc to $(i,v)$.
    Since $S$ contains at most $m$ sources, each with arcs to at most $m$ copies of $v$, we have $\sum_{i\in[m^2]}h_{i,v}\le m^2$ for every $v\in V$.
    For $z\in V$, let $R_z \coloneqq \setst{v \in V}{z \in C(v)}$ in a realization of $G$.
    Since copies are independent and use the same probabilities as $G$, the probability that no source in $S$ activates $(i,z)$ is $\E\bigl[\prod_{v\in R_z}(1-x_v)^{h_{i,v}}\bigr]$.
    For each fixed $R\subseteq V$, the AM--GM inequality gives
    \begin{equation*}
        \frac{1}{m^2}\sum_{i\in[m^2]}\prod_{v\in R}(1-x_v)^{h_{i,v}} \geq \prod_{v\in R}(1-x_v)^{m^{-2}\sum_{i\in[m^2]} h_{i,v}} \geq \prod_{v\in R}(1-x_v),
    \end{equation*}
    where we use the convention that $0^0=1$ to account for vertices $v$ with $x_v = 1$.
    The last product is the probability that $S_{\mathbf x}\cap R=\emptyset$, which, for $R=R_z$, happens exactly when $z\notin C(S_{\mathbf x})$.
    Therefore, on average over $i\in[m^2]$, the sources in $S$ activate $(i,z)$ with probability at most $\Pr(z\in C(S_{\mathbf x}))$, and summing with weights $w_z$ bounds their expected influence in $H$ by $m^2\E[\sigma_G(S_{\mathbf x})]$.
    Since a seed placed inside a copy adds at most $w(V)$, we obtain
    \begin{equation*}
        \Static_m(H) \leq m^2\,\E[\sigma_G(S_{\mathbf x})]+m\,w(V).
    \end{equation*}
    
    Dividing the two bounds and letting $m\to\infty$ gives
    \begin{equation*}
        \liminf_{m\to\infty}
        \frac{\Adaptive_m(H)}{\Static_m(H)}
        \ge \frac{\E[\sigma_G(\pi)]}{\E[\sigma_G(S_{\mathbf x})]}
    \end{equation*}
    whenever the denominator is positive. Taking the supremum over
    $G$, $k$, and $\pi\in\Pi_k$ proves the reverse inequality.
\end{proof}

\subsection{An Instance With Sampling Gap of \texorpdfstring{$1+\varphi$}{1+φ}}\label{sec:tree_phi}
In this section, we provide an asymptotic family of instances and adaptive policies whose sampling gap approaches $1+\varphi$.

\begin{lemma}\label{lem:tree_sampling_gap}
For every $\varepsilon>0$, there exist a binary tree with all arcs directed toward its root $r$, with weight one at $r$ and zero elsewhere, and an adaptive policy $\pi$ such that
\[
    \frac{\E[\sigma(\pi)]}{\E[\sigma(S_{\mathbf x(\pi)})]} \ge 1+\varphi-\varepsilon.
\]
\end{lemma}
\begin{proof}
Since only the root has positive weight, the expected influence of a policy is the probability that it activates the root.
We construct a policy that searches for a live path to the root, starting at the leaves.

\paragraph{Instance and Policy.}
Fix integers $1\le t<h$.
Take a complete binary tree of height $h$, with root $r$ at depth zero and leaves at depth $h$.
Each nonroot vertex has one arc pointing to its parent, live independently with probability $1/2$.
Give the root $r$ weight one and every other vertex weight zero.

Define the policy $\pi$ as follows.
Initially, no vertex is discarded.
For each depth $j=h,h-1,\ldots,t$, draw a uniformly random order of the vertices at depth $j$, independently of the arcs and of all previous orders.
Visit the vertices at that depth in the sampled order and apply the following steps to each vertex $v$.
\begin{enumerate}
    \item If $v$ is discarded, continue to the next vertex.
    \item Seed $v$ and observe $C(v)$.
    If $r\in C(v)$, stop the policy.
    \item Otherwise, let $u$ be the vertex of $C(v)$ closest to $r$.
    Discard $u$ and all its descendants.
\end{enumerate}
Stop after visiting every vertex at depth at least $t$.

The cascade of any vertex is always an upward path in the tree, which may or may not reach the root.
If it does not, then neither the top vertex $u$ of that path nor any of its descendants can reach the root.
Hence, $\pi$ only probes vertices whose unique path to the root has no arc whose status has been revealed.
Thus, conditional on the history, a probe at depth $j$ reaches $r$ with probability $2^{-j}$.

\paragraph{Adaptive Value and Seed Probabilities.}
Let $s_j$ be the probability that some vertex at depth $j$ has a live path to $r$.
The two subtrees below the root are independent, and a path through either child requires its arc to the root to be live.
Hence, $s_0=1$ and
\[
    s_{j+1}=1-(1-s_j/2)^2=s_j-\frac{s_j^2}{4}.
\]
The policy succeeds at the greatest depth from which a vertex can reach $r$, provided this depth is at least $t$.
Indeed, every probe at a greater depth fails, and the policy never discards a vertex that can reach $r$.
It follows that $\E[\sigma(\pi)]=s_t$.

By symmetry, the seed probabilities $x_v(\pi)$ are equal for all vertices $v$ at depth $j$, which we denote by  $x_j$.
Since each probe at depth $j$ succeeds with conditional probability $2^{-j}$ and at most one probe succeeds, the probability of success at that depth is $2^{j}\cdot 2^{-j} x_j=x_j$.
For $t\le j<h$, success at depth $j$ means that some vertex at depth $j$ reaches $r$ and no vertex at depth $j+1$ does.
Thus,
\[
    x_j=
    \begin{cases}
        0, & 0\le j<t,\\
        s_j-s_{j+1}=s_j^2/4, & t\le j<h,\\
        s_h, & j=h.
    \end{cases}
\]

\paragraph{Independent Sampling.}
Let $q_j$ be the probability that a given vertex at depth $j$ is activated by $S_{\mathbf x(\pi)}$, so that $\E[\sigma(S_{\mathbf x(\pi)})]=q_0$.
Write $f(z)=1-(1-z/2)^2=z-z^2/4$.
A vertex is activated if it is sampled or if one of its children is activated and the arc from that child is live.
Independence gives $q_h=x_h$ and
\[
    q_j=x_j+(1-x_j)f(q_{j+1}) \qquad (0\le j<h).
\]

Let $\gamma=1/\varphi$, so $\gamma^2+\gamma=1$.
We claim that $q_j\le\gamma s_{j-1}+(1-\gamma)s_h$ for every $t\le j\le h$.
At the leaves, $q_h=s_h\le\gamma s_{h-1}+(1-\gamma)s_h$.
For the induction step, suppose $q_{j+1}\le\gamma s_j+(1-\gamma)s_h$.
Since $f$ is increasing and $1$-Lipschitz on $[0,1]$, we have
\[
    q_j
    \le x_j+f(\gamma s_j)+(1-\gamma)s_h
    =\gamma\left(s_j+\frac{s_j^2}{4}\right)+(1-\gamma)s_h
    \le\gamma s_{j-1}+(1-\gamma)s_h.
\]
The equality uses $x_j=s_j^2/4$ and $1-\gamma^2=\gamma$.
The last inequality follows from $s_j^2/4\le s_{j-1}^2/4=s_{j-1}-s_j$.

No vertex at depth less than $t$ is sampled, so $q_j=f(q_{j+1})$ for $j<t$.
Since $1/f(z)-1/z=1/(4-z)\ge1/4$ for $0<z\le1$, summing over these $t$ levels gives
\[
    \frac{s_t}{q_0}
    \ge \frac{s_t}{q_t}+\frac{t s_t}{4}
    \ge \frac{s_t}{\gamma s_{t-1}+(1-\gamma)s_h}+\frac{t s_t}{4}.
\]

\paragraph{Choice of Depths.}
The recurrence for $s_j$ gives $s_j\downarrow0$ and $1/s_{j+1}-1/s_j=1/(4-s_j)\to1/4$.
Averaging these increments yields $j s_j\to4$.
Take $h=t^2$ and let $t\to\infty$.
Then $s_{t-1}/s_t\to1$ and $s_h/s_t\to0$, so the lower bound on $s_t/q_0$ tends to $1/\gamma+1=1+\varphi$.
Choosing $t$ sufficiently large proves the lemma.
\end{proof}

Since the construction in the proof of \cref{lem:sampling_gap} preserves acyclicity, \cref{lem:tree_sampling_gap} gives a worst-case adaptivity gap of at least $1+\varphi$ on directed acyclic graphs.

\section{Conclusion}
We established new upper bounds on the adaptivity gap of influence maximization under full-adoption feedback, for undirected graphs, directed graphs, and bi-directed forests. Our proofs develop two approaches. The first uses percolation inequalities to control the dependence between cascades. The second uses an expected depletion argument to show that a few rounds of static seeding leave only vertices of small expected influence, which bounds the influence of an adaptive policy. 
We showed that the worst-case sampling gap is equal to the worst-case adaptivity gap for this problem, which we used to give a lower bound of $1+\varphi$, which is above $e/(e-1)$. 
In particular, this establishes that the nature of influence maximization is different from that of submodular maximization.
We hope these ideas will also prove useful in other stochastic diffusion problems over graphs.

Several questions remain open. First, can we find analogs of the percolation inequalities for directed graphs? Some direct analogs of the tools used in this work fail in that setting.
It would be of independent interest to further extend some of this machinery to directed graphs.
Second, and most importantly, is the adaptivity gap bounded by a universal constant?
Our improved lower bound remains constant.
This raises the question of whether better upper bounds can be achieved or whether the lower bounds can be improved. 
Finally, can our methods be used to improve adaptivity-gap bounds for other diffusion models, such as the linear threshold model?

\section*{AI Methodology}
\label{sec:ai_methodology}
We developed an $\mathcal{O}(\log n)$ bound for directed graphs in \cref{sec:directed} independently and used ChatGPT to optimize it to $\mathcal{O}(\log n/\log\log n)$.
We also used ChatGPT to investigate the connection between influence maximization and recent breakthroughs on percolation inequalities and write the first draft of the proof.
We substantially rewrote that proof, and the argument presented in \cref{sec:undirected} differs considerably from the initial proposal.

We also used ChatGPT to help evaluate the mathematical statements and proofs of results from prior work used in this paper, and to revise the writing and proofs throughout the paper, and Claude to verify references and proofread the manuscript; the authors independently verified all mathematical claims and take full responsibility for all the results.

\section*{Acknowledgments}
We thank Thiago Lima Oliveira for his helpful comments on the exposition of this work.

\printbibliography

@article{Harris60,
  author = {Harris, Theodore E.},
  title = {A Lower Bound for the Critical Probability in a Certain Percolation Process},
  journal = {Proceedings of the Cambridge Philosophical Society},
  volume = {56},
  number = {1},
  pages = {13--20},
  publisher = {Cambridge University Press},
  year = {1960},
  month = jan,
  doi = {10.1017/S0305004100034241}
}

@article{Kesten80,
  author = {Kesten, Harry},
  title = {The Critical Probability of Bond Percolation on the Square Lattice Equals {$1/2$}},
  journal = {Communications in Mathematical Physics},
  volume = {74},
  number = {1},
  pages = {41--59},
  publisher = {Springer},
  year = {1980},
  month = feb,
  doi = {10.1007/BF01197577}
}

@article{HaraS90,
  author = {Hara, Takashi and Slade, Gordon},
  title = {Mean-Field Critical Behaviour for Percolation in High Dimensions},
  journal = {Communications in Mathematical Physics},
  volume = {128},
  number = {2},
  pages = {333--391},
  publisher = {Springer},
  year = {1990},
  month = mar,
  doi = {10.1007/BF02108785}
}

@book{Grimmett99,
  author = {Grimmett, Geoffrey},
  title = {Percolation},
  series = {Grundlehren der mathematischen Wissenschaften},
  volume = {321},
  edition = {2nd},
  publisher = {Springer},
  address = {Berlin},
  year = {1999},
  doi = {10.1007/978-3-662-03981-6}
}

@inproceedings{ChekuriVZ10,
  author = {Chekuri, Chandra and Vondr{\'a}k, Jan and Zenklusen, Rico},
  title = {Dependent Randomized Rounding via Exchange Properties of Combinatorial Structures},
  booktitle = {Proceedings of the 51st Annual {IEEE} Symposium on Foundations of Computer Science},
  pages = {575--584},
  publisher = {IEEE Computer Society},
  year = {2010},
  doi = {10.1109/FOCS.2010.60}
}

@article{GolovinK11,
  author = {Golovin, Daniel and Krause, Andreas},
  title = {Adaptive Submodularity: Theory and Applications in Active Learning and Stochastic Optimization},
  journal = {Journal of Artificial Intelligence Research},
  volume = {42},
  pages = {427--486},
  publisher = {AI Access Foundation},
  year = {2011},
  month = nov,
  url = {https://jair.org/index.php/jair/article/view/10731}
}

@inproceedings{GolovinK10,
  author = {Golovin, Daniel and Krause, Andreas},
  title = {Adaptive Submodularity: A New Approach to Active Learning and Stochastic Optimization},
  booktitle = {Proceedings of the 23rd Annual Conference on Learning Theory ({COLT})},
  year = {2010},
  note = {arXiv:1003.3967v2}
}

@inproceedings{KhannaL14,
  author = {Khanna, Sanjeev and Lucier, Brendan},
  title = {Influence Maximization in Undirected Networks},
  booktitle = {Proceedings of the Twenty-Fifth Annual {ACM-SIAM} Symposium on Discrete Algorithms},
  pages = {1482--1496},
  publisher = {SIAM},
  year = {2014},
  doi = {10.1137/1.9781611973402.109},
  note = {Full version: \url{https://www.cis.upenn.edu/~sanjeev/papers/soda14_undirected_influence.pdf}}
}

@inproceedings{KempeKT03,
  author = {Kempe, David and Kleinberg, Jon and Tardos, {\'E}va},
  title = {Maximizing the spread of influence through a social network},
  booktitle = {Proceedings of the Ninth {ACM SIGKDD} International Conference on Knowledge Discovery and Data Mining},
  pages = {137--146},
  publisher = {ACM},
  year = {2003},
  doi = {10.1145/956750.956769}
}

@article{AsadpourN16,
  author = {Asadpour, Arash and Nazerzadeh, Hamid},
  title = {Maximizing Stochastic Monotone Submodular Functions},
  journal = {Management Science},
  volume = {62},
  number = {8},
  pages = {2374--2391},
  publisher = {INFORMS},
  year = {2016},
  month = aug,
  doi = {10.1287/mnsc.2015.2254}
}

@inproceedings{GuptaNS17,
  author = {Gupta, Anupam and Nagarajan, Viswanath and Singla, Sahil},
  title = {Adaptivity Gaps for Stochastic Probing: Submodular and {XOS} Functions},
  booktitle = {Proceedings of the Twenty-Eighth Annual {ACM-SIAM} Symposium on Discrete Algorithms},
  pages = {1688--1702},
  publisher = {SIAM},
  year = {2017},
  doi = {10.1137/1.9781611974782.111}
}

@article{FitznerH17,
  author = {Fitzner, Robert and {van der Hofstad}, Remco},
  title = {Mean-Field Behavior for Nearest-Neighbor Percolation in {$d>10$}},
  journal = {Electronic Journal of Probability},
  volume = {22},
  pages = {1--65},
  publisher = {Institute of Mathematical Statistics},
  year = {2017},
  doi = {10.1214/17-EJP56},
  note = {Paper no. 43}
}

@inproceedings{BradacSZ19,
  author = {Bradac, Domagoj and Singla, Sahil and Zuzic, Goran},
  title = {{(Near)} Optimal Adaptivity Gaps for Stochastic Multi-Value Probing},
  editor = {Achlioptas, Dimitris and V{\'e}gh, L{\'a}szl{\'o} A.},
  booktitle = {Approximation, Randomization, and Combinatorial Optimization. Algorithms and Techniques ({APPROX/RANDOM} 2019)},
  series = {Leibniz International Proceedings in Informatics ({LIPIcs})},
  volume = {145},
  pages = {49:1--49:21},
  publisher = {Schloss Dagstuhl -- Leibniz-Zentrum f{\"u}r Informatik},
  year = {2019},
  doi = {10.4230/LIPIcs.APPROX-RANDOM.2019.49}
}

@inproceedings{PengC19,
  author = {Peng, Binghui and Chen, Wei},
  title = {Adaptive influence maximization with myopic feedback},
  editor = {Wallach, H. and Larochelle, H. and Beygelzimer, A. and d'Alch{\'e}-Buc, F. and Fox, E. and Garnett, R.},
  booktitle = {Advances in Neural Information Processing Systems},
  volume = {32},
  publisher = {Curran Associates, Inc.},
  year = {2019},
  url = {https://proceedings.neurips.cc/paper/2019/hash/f6b5f8c32c65fee991049a55dc97d1ce-Abstract.html}
}

@inproceedings{ChenP19,
  author = {Chen, Wei and Peng, Binghui},
  title = {On Adaptivity Gaps of Influence Maximization Under the Independent Cascade Model with Full-Adoption Feedback},
  editor = {Lu, Pinyan and Zhang, Guochuan},
  booktitle = {30th International Symposium on Algorithms and Computation ({ISAAC} 2019)},
  series = {Leibniz International Proceedings in Informatics ({LIPIcs})},
  volume = {149},
  pages = {24:1--24:19},
  publisher = {Schloss Dagstuhl -- Leibniz-Zentrum f{\"u}r Informatik},
  year = {2019},
  doi = {10.4230/LIPIcs.ISAAC.2019.24}
}

@article{DAngeloPV23,
  author = {D'Angelo, Gianlorenzo and Poddar, Debashmita and Vinci, Cosimo},
  title = {Better Bounds on the Adaptivity Gap of Influence Maximization under Full-Adoption Feedback},
  journal = {Artificial Intelligence},
  volume = {318},
  pages = {103895},
  publisher = {Elsevier},
  year = {2023},
  month = may,
  doi = {10.1016/j.artint.2023.103895}
}

@misc{KozmaN24,
  author = {Kozma, Gady and Nitzan, Shahaf},
  title = {A Reduction of the {$\theta(p_c)=0$} Problem to a Conjectured Inequality},
  year = {2024},
  howpublished = {arXiv:2401.12397v1 [math.PR]},
  eprint = {2401.12397v1},
  archivePrefix = {arXiv},
  primaryClass = {math.PR},
  url = {https://arxiv.org/abs/2401.12397v1}
}

@misc{Gladkov24,
  author = {Gladkov, Nikita},
  title = {Percolation Inequalities and Decision Trees},
  year = {2024},
  howpublished = {arXiv:2408.08457v2 [math.PR]},
  eprint = {2408.08457v2},
  archivePrefix = {arXiv},
  primaryClass = {math.PR},
  url = {https://arxiv.org/abs/2408.08457v2}
}

@misc{Leder26,
  author = {Leder, Justin},
  title = {{$\theta(p_c)=0$} for {Bernoulli} Bond Percolation on {$\mathbb{Z}^d$} in All Dimensions {$d\ge2$}, in {Lean 4 / Mathlib}},
  year = {2026},
  howpublished = {GitHub repository},
  url = {https://github.com/anthropics/formal-math/tree/795efb86f191735c5481675763537cfb4ff37e55/percolation},
}

@misc{BouRabee26,
  author = {Bou-Rabee, Ahmed},
  title = {Percolation after {Anthropic}: {Lean 4} Proofs of the {Kozma--Nitzan} Connection Inequalities and of Site Percolation at Criticality},
  year = {2026},
  howpublished = {GitHub repository},
  url = {https://github.com/nitromannitol/percolation-after-anthropic/tree/cd090736b8f5984d49926e80363910e7c6441a4f},
}

@misc{BlancRenaudie26,
  author = {Blanc-Renaudie, Arthur},
  title = {Notes on {Kozma} and {Nitzan}'s Proof of {$\theta(p_c)=0$}},
  year = {2026},
  howpublished = {arXiv:2610.10502v1 [math.PR]},
  eprint = {2610.10502v1},
  archivePrefix = {arXiv},
  primaryClass = {math.PR},
  url = {https://arxiv.org/abs/2610.10502v1}
}

@misc{Gouere26,
  author = {Gou{\'e}r{\'e}, Jean-Baptiste},
  title = {On the Proof of the {Kozma--Nitzan} Conjecture},
  year = {2026},
  howpublished = {arXiv:2610.30063v1 [math.PR]},
  eprint = {2610.30063v1},
  archivePrefix = {arXiv},
  primaryClass = {math.PR},
  url = {https://arxiv.org/abs/2610.30063v1}
}

@misc{OpenAI26,
  author = {{OpenAI}},
  title = {Critical Bond and Site Percolation on the Cubic Lattice},
  year = {2026},
  howpublished = {Preprint in the GitHub repository \texttt{openai/math}},
  url = {https://github.com/openai/math/blob/adc7f1241b42e322a6451854ab7e4b4c146bf78a/preprints/Critical-bond-and-site-percolation-on-the-cubic-lattice-September-24-2026/paper.pdf},
}

@article{NemhauserWF78,
  author = {Nemhauser, George L. and Wolsey, Laurence A. and Fisher, Marshall L.},
  title = {An analysis of approximations for maximizing submodular set functions---{I}},
  journal = {Mathematical Programming},
  volume = {14},
  number = {1},
  pages = {265--294},
  publisher = {Springer},
  year = {1978},
  doi = {10.1007/BF01588971}
}

@article{LiFWT18,
  author = {Li, Yuchen and Fan, Ju and Wang, Yanhao and Tan, Kian-Lee},
  title = {Influence maximization on social graphs: A survey},
  journal = {IEEE Transactions on Knowledge and Data Engineering},
  volume = {30},
  number = {10},
  pages = {1852--1872},
  publisher = {IEEE},
  year = {2018},
  doi = {10.1109/TKDE.2018.2807843}
}

@article{BanerjeeJP20,
  author = {Banerjee, Suman and Jenamani, Mamata and Pratihar, Dilip Kumar},
  title = {A survey on influence maximization in a social network},
  journal = {Knowledge and Information Systems},
  volume = {62},
  number = {9},
  pages = {3417--3455},
  publisher = {Springer},
  year = {2020},
  doi = {10.1007/s10115-020-01461-4}
}

@inproceedings{DomingosR01,
  author = {Domingos, Pedro and Richardson, Matt},
  title = {Mining the network value of customers},
  booktitle = {Proceedings of the Seventh {ACM SIGKDD} International Conference on Knowledge Discovery and Data Mining},
  pages = {57--66},
  publisher = {ACM},
  year = {2001},
  doi = {10.1145/502512.502525}
}

@inproceedings{RichardsonD02,
  author = {Richardson, Matthew and Domingos, Pedro},
  title = {Mining knowledge-sharing sites for viral marketing},
  booktitle = {Proceedings of the Eighth {ACM SIGKDD} International Conference on Knowledge Discovery and Data Mining},
  pages = {61--70},
  publisher = {ACM},
  year = {2002},
  doi = {10.1145/775047.775057}
}

@article{MosselR10,
  author = {Mossel, Elchanan and Roch, S{\'e}bastien},
  title = {Submodularity of influence in social networks: From local to global},
  journal = {SIAM Journal on Computing},
  volume = {39},
  number = {6},
  pages = {2176--2188},
  publisher = {SIAM},
  year = {2010},
  doi = {10.1137/080714452}
}

@inproceedings{ChenWW10,
  author = {Chen, Wei and Wang, Chi and Wang, Yajun},
  title = {Scalable influence maximization for prevalent viral marketing in large-scale social networks},
  booktitle = {Proceedings of the 16th {ACM SIGKDD} International Conference on Knowledge Discovery and Data Mining},
  pages = {1029--1038},
  publisher = {ACM},
  year = {2010},
  doi = {10.1145/1835804.1835934}
}

@inproceedings{BorgsBCL14,
  author = {Borgs, Christian and Brautbar, Michael and Chayes, Jennifer and Lucier, Brendan},
  title = {Maximizing social influence in nearly optimal time},
  booktitle = {Proceedings of the Twenty-Fifth Annual {ACM-SIAM} Symposium on Discrete Algorithms},
  pages = {946--957},
  publisher = {SIAM},
  year = {2014},
  doi = {10.1137/1.9781611973402.70}
}

@inproceedings{TangXS14,
  author = {Tang, Youze and Xiao, Xiaokui and Shi, Yanchen},
  title = {Influence maximization: Near-optimal time complexity meets practical efficiency},
  booktitle = {Proceedings of the 2014 {ACM SIGMOD} International Conference on Management of Data},
  pages = {75--86},
  publisher = {ACM},
  year = {2014},
  doi = {10.1145/2588555.2593670}
}

@inproceedings{DAngeloPV21,
  author = {D'Angelo, Gianlorenzo and Poddar, Debashmita and Vinci, Cosimo},
  title = {Improved Approximation Factor for Adaptive Influence Maximization via Simple Greedy Strategies},
  editor = {Bansal, Nikhil and Merelli, Emanuela and Worrell, James},
  booktitle = {48th International Colloquium on Automata, Languages, and Programming ({ICALP} 2021)},
  series = {Leibniz International Proceedings in Informatics ({LIPIcs})},
  volume = {198},
  pages = {59:1--59:19},
  publisher = {Schloss Dagstuhl -- Leibniz-Zentrum f{\"u}r Informatik},
  year = {2021},
  doi = {10.4230/LIPIcs.ICALP.2021.59}
}

@phdthesis{Vondrak07,
  author = {Vondr{\'a}k, Jan},
  title = {Submodularity in combinatorial optimization},
  school = {Univerzita Karlova, Matematicko-fyzik{\'a}ln{\'\i} fakulta},
  year = {2007},
  url = {https://dspace.cuni.cz/handle/20.500.11956/13738}
}

@incollection{GyarfasL70,
  author = {Gy{\'a}rf{\'a}s, A. and Lehel, J.},
  title = {A {Helly}-type problem in trees},
  editor = {Erd{\H{o}}s, P. and R{\'e}nyi, A. and S{\'o}s, V. T.},
  booktitle = {Combinatorial Theory and its Applications},
  series = {Colloquia Mathematica Societatis J{\'a}nos Bolyai},
  volume = {4},
  pages = {571--584},
  publisher = {North-Holland},
  address = {Amsterdam},
  year = {1970},
  url = {https://www.renyi.hu/~gyarfas/Cikkek/02_GyarfasLehel_AHellyTypeProblemInTrees.pdf}
}

@inproceedings{mossel2007submodularity,
  title={On the submodularity of influence in social networks},
  author={Mossel, Elchanan and Roch, S{\'e}bastien},
  booktitle={Proceedings of the Thirty-Ninth Annual {ACM} Symposium on Theory of Computing},
  pages={128--134},
  year={2007}
}

@inproceedings{DeanGV04,
  author    = {Brian C. Dean and Michel X. Goemans and Jan Vondr{\'a}k},
  title     = {Approximating the Stochastic Knapsack Problem:
               The Benefit of Adaptivity},
  booktitle = {Proceedings of the 45th Annual IEEE Symposium on
               Foundations of Computer Science (FOCS)},
  pages     = {208--217},
  year      = {2004},
  publisher = {IEEE Computer Society},
  doi       = {10.1109/FOCS.2004.15}
}

@inproceedings{li2025adaptivity,
  title={Adaptivity gaps for stochastic probing with subadditive functions},
  author={Li, Jian and Liu, Yinchen and Zhang, Yiran},
  booktitle={2025 IEEE 66th Annual Symposium on Foundations of Computer Science (FOCS)},
  pages={149--185},
  year={2025},
  organization={IEEE}
}

@article{jankowski2017balancing,
  title={Balancing speed and coverage by sequential seeding in complex networks},
  author={Jankowski, Jaros{\l}aw and Br{\'o}dka, Piotr and Kazienko, Przemys{\l}aw and Szyma{\'n}ski, Boles{\l}aw K. and Michalski, Rados{\l}aw and Kajdanowicz, Tomasz},
  journal={Scientific Reports},
  volume={7},
  number={1},
  pages={891},
  year={2017},
  publisher={Nature Publishing Group UK, London}
}

@article{banerjee2013diffusion,
  title={The diffusion of microfinance},
  author={Banerjee, Abhijit and Chandrasekhar, Arun G. and Duflo, Esther and Jackson, Matthew O.},
  journal={Science},
  volume={341},
  number={6144},
  pages={1236498},
  year={2013},
  publisher={American Association for the Advancement of Science}
}

@article{hinz2011seeding,
  title={Seeding strategies for viral marketing: An empirical comparison},
  author={Hinz, Oliver and Skiera, Bernd and Barrot, Christian and Becker, Jan U.},
  journal={Journal of Marketing},
  volume={75},
  number={6},
  pages={55--71},
  year={2011},
  publisher={SAGE Publications, Sage CA: Los Angeles, CA}
}

@inproceedings{leskovec2007cost,
  title={Cost-effective outbreak detection in networks},
  author={Leskovec, Jure and Krause, Andreas and Guestrin, Carlos and Faloutsos, Christos and VanBriesen, Jeanne and Glance, Natalie},
  booktitle={Proceedings of the 13th {ACM SIGKDD} International Conference on Knowledge Discovery and Data Mining},
  pages={420--429},
  year={2007}
}

@inproceedings{yadav2016using,
  title={Using Social Networks to Aid Homeless Shelters: Dynamic Influence Maximization under Uncertainty},
  author={Yadav, Amulya and Chan, Hau and Jiang, Albert Xin and Xu, Haifeng and Rice, Eric and Tambe, Milind},
  booktitle={Proceedings of the 2016 International Conference on Autonomous Agents \& Multiagent Systems},
  pages={740--748},
  year={2016}
}

@article{agrawal2012price,
  title={Price of correlations in stochastic optimization},
  author={Agrawal, Shipra and Ding, Yichuan and Saberi, Amin and Ye, Yinyu},
  journal={Operations Research},
  volume={60},
  number={1},
  pages={150--162},
  year={2012},
  publisher={INFORMS}
}

@article{el2026two,
  title={Two-sided assortment optimization: Adaptivity gaps and approximation algorithms},
  author={El Housni, Omar and Hennebelle, Ulysse and Torrico, Alfredo},
  journal={Operations Research},
  year={2026},
  publisher={INFORMS}
}

@article{bansal2015adaptivity,
  title={On the adaptivity gap of stochastic orienteering},
  author={Bansal, Nikhil and Nagarajan, Viswanath},
  journal={Mathematical Programming},
  volume={154},
  number={1},
  pages={145--172},
  year={2015},
  publisher={Springer}
}

@inproceedings{jiang2020stochastic,
  author    = {Jiang, Haotian and Li, Jian and Liu, Daogao and Singla, Sahil},
  title     = {Algorithms and Adaptivity Gaps for Stochastic {$k$-TSP}},
  booktitle = {11th Innovations in Theoretical Computer Science Conference (ITCS 2020)},
  volume    = {151},
  pages     = {45:1--45:25},
  year      = {2020}
}

\end{document}